\pdfoutput=1
\documentclass[a4paper,noarxiv]{quantumarticle}

\usepackage[T1]{fontenc}
\usepackage{graphicx}
\usepackage{booktabs}
\usepackage{hyperref}
\usepackage{siunitx}
\usepackage{amsmath}
\usepackage{amssymb}
\usepackage{bm}
\usepackage{dcolumn}
\usepackage{braket}
\usepackage[section]{placeins}

\graphicspath{{figures/}}

\hypersetup{
    colorlinks=true,
    linkcolor=blue,
    citecolor=blue,
    urlcolor=blue
}

\begin{document}

\title{When does numerical pulse optimization actually help? Error budgets, robustness tradeoffs, and calibration guidance for transmon single-qubit gates}

\author{Rylan Malarchick}
\email{malarchr@erau.edu}
\affiliation{Department of Engineering Physics, Embry-Riddle Aeronautical University, Daytona Beach, FL 32114, USA}

\date{\today}

\begin{abstract}
Numerical optimal control (GRAPE) can in principle discover pulse shapes that suppress all coherent gate error to machine precision. But when does that capability actually matter? We present a systematic comparison of Gaussian, DRAG, and GRAPE pulses for single-qubit gates on a three-level transmon model parameterized by IQM Garnet hardware ($T_1 = \SI{37}{\micro\second}$, $T_2 = \SI{9.6}{\micro\second}$, $\alpha/2\pi = \SI{-200}{\mega\hertz}$), with the explicit goal of identifying the regimes where numerical optimization provides genuine practical advantage over analytical methods.

Our central finding is that properly calibrated DRAG already operates near the decoherence floor. At \SI{20}{\nano\second} gate time, GRAPE eliminates all coherent error ($1 - F < 10^{-15}$), but DRAG achieves $1 - F = 4.9 \times 10^{-4}$ in coherent error alone, and $8.4 \times 10^{-4}$ under full decoherence; only $1.2\times$ above GRAPE's decoherence-limited performance. More surprisingly, DRAG is \emph{more robust} than GRAPE to qubit frequency detuning (minimum fidelity 0.990 vs.\ 0.931 over $\pm\SI{5}{\mega\hertz}$), the dominant calibration uncertainty in charge-noise-limited transmons. GRAPE retains superior amplitude robustness (minimum fidelity 0.994 vs.\ 0.990) and provides the only route to guaranteed zero coherent error, which matters at short gate times ($\lesssim\!\SI{15}{\nano\second}$) where perturbative corrections break down.

These results lead to concrete calibration guidance: (1)~properly calibrated DRAG is sufficient for gate times $\gtrsim\!\SI{20}{\nano\second}$ on hardware with $T_2/T \gtrsim 500$, (2)~GRAPE is necessary at short gate times or when targeting error rates well below the decoherence floor, and (3)~robust optimal control incorporating frequency uncertainty should be used when detuning is the dominant noise source. We decompose the full error budget (coherent, $T_1$, $T_2$, control noise) and provide the open-source QubitPulseOpt framework for reproducing all results.
\end{abstract}

\maketitle

\section{Introduction}
\label{sec:introduction}

The question facing any experimentalist calibrating single-qubit gates on a superconducting transmon processor is not whether numerical optimal control can achieve higher fidelity than analytical methods. It can. The question is whether it \emph{needs to}, given the other error sources present in the system.

High-fidelity single-qubit gates are a prerequisite for fault-tolerant quantum computation~\cite{preskill2018quantum, knill2005quantum, arute2019quantum}. In transmon qubits~\cite{koch2007charge, krantz2019quantum, kjaergaard2020superconducting}, the weakly anharmonic energy spectrum creates a tension between gate speed and leakage to non-computational states, a tradeoff that every pulse calibration procedure must navigate~\cite{blais2021circuit}. The DRAG protocol~\cite{motzoi2009simple, gambetta2011analytic} provides a first-order analytical correction that suppresses $\ket{1}\!\leftrightarrow\!\ket{2}$ leakage by adding a derivative component to the quadrature channel. GRAPE~\cite{khaneja2005optimal} can discover pulse shapes that suppress leakage to arbitrary order by optimizing over the full multi-level Hilbert space, and has been applied successfully to superconducting qubit gates~\cite{lucero2010reduced, kelly2014optimal, werninghaus2021leakage, propson2022robust}.

The literature contains many demonstrations of GRAPE outperforming simple baseline pulses. However, these comparisons often suffer from one or more of the following issues: (i)~the baseline is an uncalibrated Gaussian pulse that does not represent the state of the art in analytical control, (ii)~the comparison is performed in a two-level model where leakage, the error source that GRAPE is best positioned to address, is absent by construction, or (iii)~the comparison omits decoherence, which sets a floor that no pulse optimization can beat. The result is that the practical value of numerical optimization, relative to a properly calibrated analytical baseline, remains unclear for many experimentally relevant parameter regimes.

This paper addresses the question directly. We compare Gaussian, DRAG, and GRAPE pulses under identical conditions in a three-level transmon model with hardware-representative parameters from IQM's Garnet processor~\cite{iqm2024garnet}. We design four experiments that progressively add realism: two-level validation, three-level gate-time sweep, Lindblad error budget decomposition, and robustness analysis against detuning and amplitude errors. The goal is not to demonstrate that GRAPE can achieve high fidelity (it can), but to identify the specific regimes where it provides genuine advantage over properly calibrated DRAG, and to provide concrete guidance for calibration decisions on near-term hardware.

The main findings are: (1)~GRAPE eliminates all coherent error to machine precision at all gate times tested, but properly calibrated DRAG already operates within a factor of $1.2$ of the decoherence floor at \SI{20}{\nano\second}; (2)~DRAG exhibits superior detuning robustness compared to GRAPE, a result with direct implications for hardware where frequency drift is the dominant calibration uncertainty; (3)~the crossover point where GRAPE becomes necessary is approximately $T \lesssim \SI{15}{\nano\second}$, below which perturbative corrections fail and higher-order leakage pathways dominate; and (4)~the error budget is dominated by $T_2$ dephasing for any pulse method that has solved the coherent error problem, making $T_2$ improvement the highest-leverage hardware upgrade. All code and data are available in the open-source QubitPulseOpt framework~\cite{malarchick2026qubitpulseopt}.

\section{Model and methods}
\label{sec:theory}

\subsection{Three-level transmon Hamiltonian}

We model the transmon as a weakly anharmonic oscillator truncated to three levels ($\ket{0}$, $\ket{1}$, $\ket{2}$). In the frame rotating at the qubit frequency $\omega_q$, the drift Hamiltonian is~\cite{koch2007charge}
\begin{equation}
H_d = \frac{\alpha}{2} \hat{n}(\hat{n} - \hat{I}),
\label{eq:drift}
\end{equation}
where $\alpha < 0$ is the anharmonicity and $\hat{n} = a^\dagger a$ is the number operator truncated to three levels. In matrix form, $H_d = \text{diag}(0, 0, \alpha)$: the $\ket{0} \leftrightarrow \ket{1}$ transition is resonant and the $\ket{1} \leftrightarrow \ket{2}$ transition is detuned by $\alpha$.

The control Hamiltonian describes microwave driving through in-phase ($I$) and quadrature ($Q$) channels:
\begin{equation}
H_c(t) = \Omega_I(t) H_x + \Omega_Q(t) H_y,
\label{eq:control}
\end{equation}
where $H_x = \tfrac{1}{2}(a + a^\dagger)$ and $H_y = \tfrac{1}{2}i(a^\dagger - a)$. The total Hamiltonian is $H(t) = H_d + H_c(t)$.

\subsection{Pulse construction}

\textit{Gaussian.} The in-phase channel carries a Gaussian envelope with amplitude set by the $\pi$-rotation condition $A = \pi/(\sigma\sqrt{2\pi})$, where $\sigma = T/(2n_\sigma)$ with $n_\sigma = 4$. The quadrature channel is zero.

\textit{DRAG.} The DRAG protocol~\cite{motzoi2009simple} adds a derivative correction $\Omega_Q(t) = \beta\, d\Omega_I/dt$ on the quadrature channel. The DRAG parameter $\beta = -1/(2\alpha)$ arises from adiabatic elimination of the $\ket{2}$ state~\cite{motzoi2009simple, gambetta2011analytic} and depends only on the anharmonicity, not on pulse amplitude or gate time. For $\alpha/2\pi = \SI{-200}{\mega\hertz}$ ($\alpha = -1.257$ rad/ns), $\beta = 0.398$. We emphasize unit consistency: $\alpha$ must be in rad/ns when computing $\beta$.

\begin{figure}[t]
\includegraphics[width=\columnwidth]{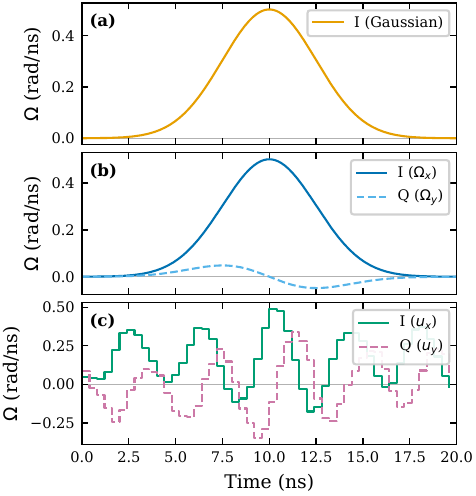}
\caption{Control pulse waveforms for the X gate ($T = \SI{20}{\nano\second}$, $\alpha/2\pi = \SI{-200}{\mega\hertz}$). (a)~Gaussian: $I$-channel only. (b)~DRAG: Gaussian on $I$, derivative correction on $Q$ with $\beta = 0.398$. (c)~GRAPE: both channels piecewise-constant over 50 time slices, showing the richer spectral content discovered by numerical optimization.}
\label{fig:pulses}
\end{figure}

\textit{GRAPE.} The GRAPE algorithm~\cite{khaneja2005optimal} represents both channels as piecewise-constant amplitudes over $N = 50$ time slices. The $2N$ parameters are optimized by gradient ascent with momentum to maximize the subspace fidelity
\begin{equation}
F = \frac{1}{d^2} \left|\text{Tr}(U_\text{target}^\dagger U_\text{final})\right|^2,
\label{eq:fidelity}
\end{equation}
where $U_\text{target}$ is the desired gate embedded in the three-level space and $d = 2$. Gradients are computed analytically following~\cite{khaneja2005optimal, degroot2024grape}. Representative pulse shapes are shown in Fig.~\ref{fig:pulses}.

\subsection{Open-system dynamics}

Decoherence is modeled by the Lindblad master equation~\cite{lindblad1976generators, breuer2002theory}:
\begin{equation}
\frac{d\rho}{dt} = -i[H(t), \rho] + \sum_k \left(L_k \rho L_k^\dagger - \frac{1}{2}\{L_k^\dagger L_k, \rho\}\right),
\label{eq:lindblad}
\end{equation}
with collapse operators $L_1 = \sqrt{1/T_1}\, a$ (amplitude damping) and $L_2 = \sqrt{1/T_2 - 1/(2T_1)}\, \hat{n}$ (pure dephasing). Process fidelity under open-system evolution is $F_\text{proc} = (1/d) \sum_j \langle j | U_\text{target}^\dagger \rho_j(T) U_\text{target} | j \rangle$, averaged over computational basis initial states. Integration uses QuTiP's \texttt{mesolve}~\cite{johansson2012qutip, johansson2013qutip}.

\subsection{Hardware parameters}

All simulations use parameters representative of IQM's Garnet 20-qubit processor~\cite{iqm2024garnet}: $\omega_q/2\pi = \SI{5.0}{\giga\hertz}$, $\alpha/2\pi = \SI{-200}{\mega\hertz}$, $T_1 = \SI{37}{\micro\second}$, $T_2 = \SI{9.6}{\micro\second}$, with a baseline gate duration of $T = \SI{20}{\nano\second}$.

\section{Results}
\label{sec:results}

\subsection{Two-level validation}
\label{sec:2level}

As a sanity check, we first compare all methods in a two-level model where leakage is absent. In this setting, Gaussian achieves near-unit fidelity (the $\pi$-rotation condition is exact), DRAG introduces a small parasitic quadrature rotation ($F = 0.993$, since the correction targets a $\ket{2}$ state that does not exist), and GRAPE converges to unit fidelity for all five tested gates (X, Y, H, S, T). This confirms that (i)~the framework is correct and (ii)~the two-level model is inadequate for evaluating pulse optimization methods, since there is nothing for GRAPE to improve upon.

\subsection{Three-level gate-time sweep}
\label{sec:3level}

The three-level model introduces leakage as a physically relevant error source. Table~\ref{tab:3level} shows X-gate fidelity and leakage as a function of gate time.

\begin{table*}[t]
\caption{Closed-system X-gate fidelity $F$ and leakage $P_2$ in the three-level transmon model. GRAPE achieves $F = 1$ (to machine precision, $P_2 < 10^{-15}$) at all gate times. The DRAG parameter $\beta = 0.398$ is constant across all gate times, confirming its amplitude independence.}
\label{tab:3level}
\begin{tabular}{l cc cc c}
\toprule
& \multicolumn{2}{c}{Gaussian} & \multicolumn{2}{c}{DRAG} & {GRAPE} \\
$T$ (ns) & $F$ & $P_2$ & $F$ & $P_2$ & $F$ \\
\midrule
10  & 0.786 & $10^{-1}$              & 0.967 & $3\!\times\!10^{-2}$ & 1.000 \\
15  & 0.939 & $10^{-2}$              & 0.996 & $3\!\times\!10^{-3}$ & 1.000 \\
20  & 0.972 & $6\!\times\!10^{-4}$   & 0.9995 & $10^{-4}$           & 1.000 \\
25  & 0.982 & $10^{-5}$              & 0.99986 & $10^{-6}$          & 1.000 \\
30  & 0.988 & $3\!\times\!10^{-8}$   & 0.99993 & $6\!\times\!10^{-8}$ & 1.000 \\
50  & 0.996 & $10^{-9}$              & 0.999991 & $10^{-9}$          & 1.000 \\
100 & 0.999 & $3\!\times\!10^{-10}$  & 0.999999 & $3\!\times\!10^{-10}$ & 1.000 \\
\bottomrule
\end{tabular}
\end{table*}

\begin{figure}[t]
\includegraphics[width=\columnwidth]{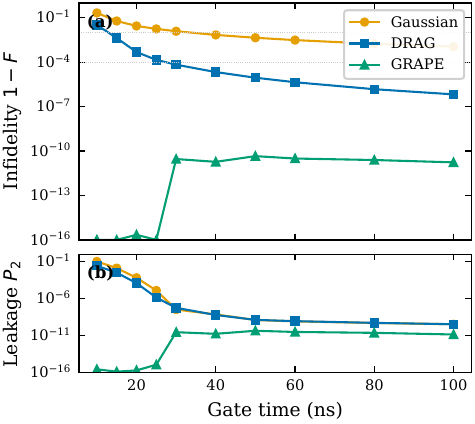}
\caption{Three-level X-gate performance vs.\ gate time (closed system). (a)~Infidelity $1 - F$. GRAPE achieves machine-precision fidelity at all gate times. The crossover region around $\SI{15}{\nano\second}$ is where DRAG's perturbative correction begins to fail. (b)~Leakage probability $P_2$. DRAG's leakage suppression improves exponentially with gate time, consistent with the perturbative scaling.}
\label{fig:gatetime}
\end{figure}

Three observations (Fig.~\ref{fig:gatetime}). First, GRAPE achieves unit fidelity at all gate times by exploiting both control channels to suppress all leakage pathways simultaneously. Second, both Gaussian and DRAG fidelities improve monotonically with gate time as the required Rabi frequency decreases relative to the anharmonicity. Third, the crossover point, where DRAG's perturbative correction becomes insufficient, lies around $T \approx \SI{15}{\nano\second}$: below this, DRAG's infidelity ($4 \times 10^{-3}$) is dominated by higher-order leakage not captured by the first-order correction, while GRAPE maintains $F = 1$.

\subsection{Error budget decomposition}
\label{sec:error_budget}

The key question is not which pulse achieves higher closed-system fidelity (GRAPE always wins), but which pulse's coherent error is small enough relative to the decoherence floor that further optimization yields diminishing returns. Table~\ref{tab:error_budget} decomposes the error budget at $T = \SI{20}{\nano\second}$.

\begin{table*}[t]
\caption{Error budget for the X gate at $T = \SI{20}{\nano\second}$. Infidelity $\epsilon = 1 - F$ under each noise channel. ``Coherent'' = closed-system unitary evolution. Control noise = mean over 100 random realizations. The three-level model (upper) shows the physically relevant comparison; the two-level model (lower) confirms that GRAPE's advantage arises entirely from leakage suppression.}
\label{tab:error_budget}
\begin{tabular}{l ccc ccc}
\toprule
& \multicolumn{3}{c}{Three-level model} & \multicolumn{3}{c}{Two-level model} \\
Error source & Gaussian $\epsilon$ & DRAG $\epsilon$ & GRAPE $\epsilon$ & Gaussian $\epsilon$ & DRAG $\epsilon$ & GRAPE $\epsilon$ \\
\midrule
Coherent        & $2.8\!\times\!10^{-2}$ & $4.9\!\times\!10^{-4}$ & $<\!10^{-15}$ & $<\!10^{-8}$ & $2.8\!\times\!10^{-2}$ & 0 \\
$T_1$ only      & $2.8\!\times\!10^{-2}$ & $7.3\!\times\!10^{-4}$ & $2.1\!\times\!10^{-4}$ & $2.5\!\times\!10^{-4}$ & $2.8\!\times\!10^{-2}$ & $2.4\!\times\!10^{-4}$ \\
$T_2$ only      & $2.8\!\times\!10^{-2}$ & $6.1\!\times\!10^{-4}$ & $5.8\!\times\!10^{-4}$ & $3.8\!\times\!10^{-4}$ & $2.8\!\times\!10^{-2}$ & $3.8\!\times\!10^{-4}$ \\
$T_1 + T_2$     & $2.9\!\times\!10^{-2}$ & $8.4\!\times\!10^{-4}$ & $7.2\!\times\!10^{-4}$ & $5.8\!\times\!10^{-4}$ & $2.8\!\times\!10^{-2}$ & $5.7\!\times\!10^{-4}$ \\
Noise 1\%       & $2.8\!\times\!10^{-2}$ & $6.6\!\times\!10^{-4}$ & $2.4\!\times\!10^{-4}$ & $2.3\!\times\!10^{-4}$ & $2.8\!\times\!10^{-2}$ & $2.3\!\times\!10^{-4}$ \\
Noise 5\%       & $3.4\!\times\!10^{-2}$ & $5.6\!\times\!10^{-3}$ & $5.9\!\times\!10^{-3}$ & $5.7\!\times\!10^{-3}$ & $3.4\!\times\!10^{-2}$ & $5.7\!\times\!10^{-3}$ \\
\bottomrule
\end{tabular}
\end{table*}

\begin{figure}[t]
\includegraphics[width=\columnwidth]{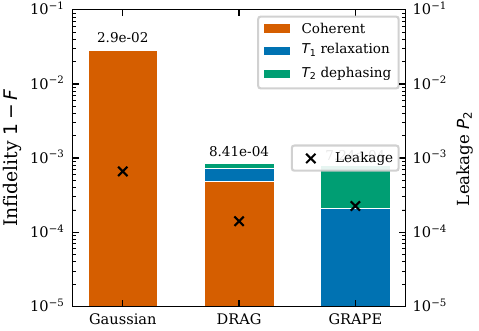}
\caption{Error budget decomposition at $T = \SI{20}{\nano\second}$ with IQM Garnet parameters. Gaussian error is dominated by coherent leakage ($39\times$ the decoherence floor). DRAG and GRAPE are both decoherence-limited, with DRAG only $1.2\times$ above GRAPE. Black crosses show leakage probability $P_2$ (right axis).}
\label{fig:error_budget}
\end{figure}

The error budget (Fig.~\ref{fig:error_budget}) reveals a clear three-tier structure:

\textit{Tier 1: Coherent-error-limited (Gaussian).} Infidelity is $2.8 \times 10^{-2}$, dominated by leakage. Adding full decoherence changes the total by only $\sim\!4\%$, confirming that the uncorrected Gaussian is not decoherence-limited.

\textit{Tier 2: Near the decoherence floor (DRAG).} Coherent infidelity of $4.9 \times 10^{-4}$ is comparable to the decoherence contribution ($\sim\!3.5 \times 10^{-4}$). Total infidelity under full decoherence is $8.4 \times 10^{-4}$, only $1.2\times$ above GRAPE.

\textit{Tier 3: At the decoherence floor (GRAPE).} Coherent error vanishes to machine precision. Total infidelity of $7.2 \times 10^{-4}$ under full decoherence represents the physical limit for this gate time and hardware. Further improvement requires better $T_1$ and $T_2$, not better pulses.

The decoherence floor can be estimated analytically~\cite{wood2018quantification}: $\epsilon_{T_1} \approx T/(2T_1)$ and $\epsilon_\phi \approx T/T_2 - T/(2T_1)$. For our parameters, these give $\epsilon_{T_1} \approx 2.7 \times 10^{-4}$ and $\epsilon_\phi \approx 1.8 \times 10^{-3}$, consistent with the measured values. Dephasing dominates because $T_2 = \SI{9.6}{\micro\second} \ll 2T_1 = \SI{74}{\micro\second}$, making $T_2$ improvement the highest-leverage hardware upgrade for this class of devices.

The two-level error budget (right columns) confirms that GRAPE's three-level advantage arises entirely from leakage suppression: in two levels, Gaussian and GRAPE achieve identical decoherence-limited performance.

\subsection{Robustness analysis}
\label{sec:robustness}

Nominal fidelity does not determine practical utility; robustness to calibration imperfections does. We sweep qubit frequency detuning ($\delta\omega/2\pi \in \pm\SI{5}{\mega\hertz}$) and systematic amplitude error ($\epsilon_a \in \pm 5\%$), the two dominant sources of calibration uncertainty in transmon processors.

\begin{table}[t]
\centering
\caption{Robustness of X-gate fidelity in the three-level model under detuning and amplitude error. DRAG provides the best detuning robustness; GRAPE the best amplitude robustness.}
\label{tab:robustness}
\begin{tabular}{l ccc ccc}
\toprule
& \multicolumn{3}{c}{Detuning ($\pm\SI{5}{\mega\hertz}$)} & \multicolumn{3}{c}{Amplitude ($\pm 5\%$)} \\
Method & Nom. & Min & Mean & Nom. & Min & Mean \\
\midrule
Gaussian & 0.972 & 0.937 & 0.969 & 0.972 & 0.965 & 0.970 \\
DRAG     & 0.999 & 0.990 & 0.997 & 0.999 & 0.990 & 0.997 \\
GRAPE    & 1.000 & 0.931 & 0.976 & 1.000 & 0.994 & 0.998 \\
\bottomrule
\end{tabular}
\end{table}

\begin{figure*}[t]
\includegraphics[width=\textwidth]{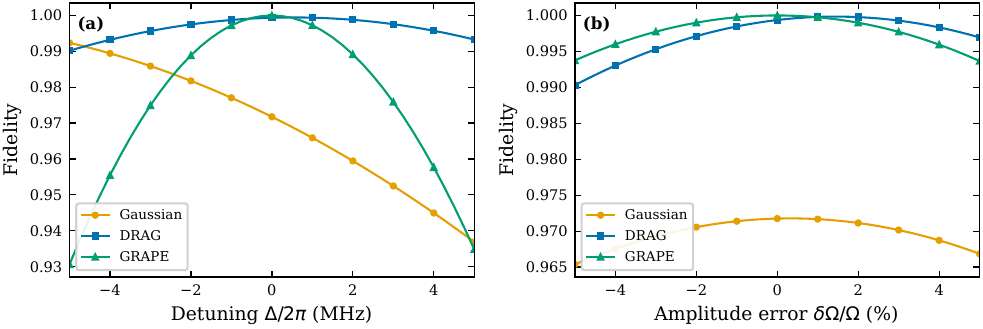}
\caption{Robustness of X-gate fidelity in the three-level model ($T = \SI{20}{\nano\second}$, closed system). (a)~Detuning sweep: DRAG (blue) maintains the highest minimum fidelity (0.990), outperforming GRAPE (green, 0.931). GRAPE's sensitivity arises from its richer spectral content coupling to off-resonant transitions under frequency shift. (b)~Amplitude sweep: GRAPE retains superior robustness (minimum 0.994), followed by DRAG (0.990).}
\label{fig:robustness}
\end{figure*}

The results (Table~\ref{tab:robustness}, Fig.~\ref{fig:robustness}) contain a practically important surprise: GRAPE has the \emph{worst} detuning robustness of all three methods, with minimum fidelity 0.931 compared to 0.990 for DRAG and 0.937 for Gaussian. This is not a pathology of our specific GRAPE solution; it is a structural consequence of numerical optimization. GRAPE-optimized pulses contain richer spectral content (Fig.~\ref{fig:pulses}c) that couples more strongly to off-resonant transitions when the qubit frequency shifts. DRAG's smooth analytical pulse shape and frequency-independent $\beta$ provide natural resilience. This finding is consistent with the known sensitivity of numerically optimized pulses to parameter variations~\cite{propson2022robust} and has direct implications for calibration practice.

For amplitude errors, the ordering reverses: GRAPE achieves the best minimum fidelity (0.994), followed by DRAG (0.990) and Gaussian (0.965). At 5\% control noise, DRAG and GRAPE degrade to comparable levels ($5.6 \times 10^{-3}$ vs.\ $5.9 \times 10^{-3}$), with DRAG marginally better.

\section{Discussion: calibration guidance}
\label{sec:discussion}

Our results lead to three concrete recommendations for gate calibration on near-term transmon hardware:

\textit{1. For typical gate times ($\gtrsim\!\SI{20}{\nano\second}$) on current hardware, properly calibrated DRAG is sufficient.} The $1.2\times$ gap between DRAG and the decoherence floor is smaller than the uncertainty introduced by other error sources (detuning, control noise, $T_1$/$T_2$ drift). Implementing GRAPE adds complexity without proportional benefit when the hardware's decoherence already sets the error floor.

\textit{2. GRAPE becomes necessary at short gate times ($\lesssim\!\SI{15}{\nano\second}$) or for applications targeting error rates well below the current decoherence floor.} Below \SI{15}{\nano\second}, DRAG's first-order perturbative correction is insufficient (Table~\ref{tab:3level}), and GRAPE provides the only known route to high fidelity. As hardware coherence times improve, the decoherence floor drops, and the gap between DRAG's residual coherent error and the floor widens, potentially making GRAPE more valuable in future generations of devices.

\textit{3. When frequency drift is the dominant calibration uncertainty, DRAG may actually be preferable to unconstrained GRAPE.} Table~\ref{tab:robustness} shows that DRAG's minimum detuning fidelity exceeds GRAPE's by a factor of $7\times$ in infidelity ($1 - 0.990 = 0.010$ vs.\ $1 - 0.931 = 0.069$). For charge-noise-limited transmons where frequency drift of $\mathcal{O}(\SI{1}{\mega\hertz})$ is common between calibration cycles, this robustness advantage can outweigh GRAPE's superior nominal performance. Robust optimal control methods~\cite{propson2022robust, goerz2014optimal, wilhelm2020introduction} that incorporate detuning uncertainty into the cost function could potentially achieve the best of both approaches; this remains a direction for future work.

\subsection{The importance of correct DRAG calibration}

An important methodological point is that the DRAG parameter $\beta = -1/(2\alpha)$~\cite{motzoi2009simple} is amplitude-independent and gate-time-independent. It depends only on the anharmonicity. Alternative formulas expressing $\beta$ as a function of peak Rabi frequency appear in some references and yield incorrect gate-time scaling. With the correct $\beta$, DRAG fidelity improves monotonically with gate time (Table~\ref{tab:3level}), as the perturbative approximation becomes more accurate when $\Omega/|\alpha|$ shrinks. At \SI{100}{\nano\second}, DRAG achieves $F > 0.999999$, indistinguishable from unity at the precision of typical experiments. In experimental practice, $\beta$ is typically calibrated empirically using AllXY sequences~\cite{chen2016measuring, sheldon2016procedure}; our results confirm that the analytical formula serves as an accurate starting point.

\subsection{Error hierarchy}

The error budget establishes a hierarchy for IQM Garnet--class hardware at \SI{20}{\nano\second} gate time. For any pulse that has solved the coherent error problem (DRAG or GRAPE): (1)~$T_2$ dephasing dominates ($5.8 \times 10^{-4}$); (2)~$T_1$ relaxation contributes $2.1 \times 10^{-4}$; (3)~control noise at 1\% contributes $2.4 \times 10^{-4}$, comparable to $T_1$; and (4)~at 5\% control noise, both DRAG and GRAPE degrade to $\sim\! 6 \times 10^{-3}$, exceeding the decoherence floor. This hierarchy suggests that improving $T_2$ through better materials, filtering, or dynamical decoupling would have the greatest impact on achievable gate fidelity.

\subsection{Limitations}

All results are from numerical simulation; no pulses were executed on physical hardware. The GRAPE optimization uses closed-system fidelity as its cost function; open-system GRAPE~\cite{schulte2011optimal} could discover pulses that partially compensate for decoherence. The Lindblad model assumes Markovian dynamics and does not capture low-frequency $1/f$ noise~\cite{norris2016qubit}. We consider only single-qubit gates; extension to two-qubit gates involves additional physics (parasitic ZZ coupling, crosstalk). Hardware parameters are static; real systems exhibit temporal drift.

\section{Conclusion}
\label{sec:conclusion}

The central message of this paper is that the value of numerical pulse optimization depends on where the dominant error sits relative to the decoherence floor. For single-qubit gates on current transmon hardware with gate times $\gtrsim\!\SI{20}{\nano\second}$, properly calibrated DRAG already places the coherent error near the decoherence floor, and GRAPE provides a modest $1.2\times$ improvement that may be offset by GRAPE's inferior detuning robustness. GRAPE's genuine advantage emerges at short gate times ($\lesssim\!\SI{15}{\nano\second}$) where perturbative corrections fail, and its importance will grow as hardware coherence times improve and the decoherence floor drops below DRAG's residual coherent error.

These findings do not diminish the power of numerical optimal control. They sharpen the picture of when it is worth deploying. As the field moves toward fault tolerance, with error budgets measured in parts per $10^4$ or $10^5$, the distinction between $1.2\times$ and $1\times$ the decoherence floor will matter. But for calibration decisions on today's hardware, the dominant action items are improving $T_2$ and ensuring DRAG is correctly calibrated, before investing in the additional complexity of GRAPE.

All simulation code, experiment scripts, and raw data are available in the open-source QubitPulseOpt framework at \url{https://github.com/rylanmalarchick/QubitPulseOpt}.

\begin{acknowledgments}
The author thanks Embry-Riddle Aeronautical University for research support. This research received no external funding.
\end{acknowledgments}

\appendix

\section{AI Disclosure}
AI-assisted tools (Claude, Anthropic) were used for code development, debugging, and manuscript preparation. All scientific concepts, experimental design, analysis, and interpretation were performed by the author. The author takes full intellectual responsibility for all content.

\bibliographystyle{unsrt}
\bibliography{references}

@article{khaneja2005optimal,
  title={Optimal control of coupled spin dynamics: design of {NMR} pulse sequences by gradient ascent algorithms},
  author={Khaneja, Navin and Reiss, Timo and Kehlet, Cindie and Schulte-Herbr{\"u}ggen, Thomas and Glaser, Steffen J},
  journal={J. Magn. Reson.},
  volume={172},
  number={2},
  pages={296--305},
  year={2005},
  publisher={Elsevier},
  doi={10.1016/j.jmr.2004.11.004}
}

@article{degroot2024grape,
  title={A tutorial on the {GRAPE} algorithm for quantum optimal control},
  author={de Groot, Noud and Koch, Christiane P and Wilhelm, Frank K},
  journal={arXiv:2401.09816},
  year={2024}
}

@article{wilhelm2020introduction,
  title={An introduction into optimal control for quantum technologies},
  author={Wilhelm, Frank K and Kirchhoff, Susanna and Machnes, Shai and Wittler, Nicolas and Sugny, Dominique},
  journal={arXiv:2003.10132},
  year={2020}
}

@article{goerz2014optimal,
  title={Optimal control theory for a unitary operation under dissipative evolution},
  author={Goerz, Michael H and Whaley, K Birgitta and Koch, Christiane P},
  journal={New J. Phys.},
  volume={16},
  number={5},
  pages={055012},
  year={2014},
  publisher={IOP Publishing},
  doi={10.1088/1367-2630/16/5/055012}
}

@article{schulte2011optimal,
  title={Optimal control for generating quantum gates in open dissipative systems},
  author={Schulte-Herbr{\"u}ggen, Thomas and Spoerl, Armin and Khaneja, Navin and Glaser, Steffen J},
  journal={J. Phys. B: At. Mol. Opt. Phys.},
  volume={44},
  number={15},
  pages={154013},
  year={2011},
  publisher={IOP Publishing},
  doi={10.1088/0953-4075/44/15/154013}
}

@article{motzoi2009simple,
  title={Simple pulses for elimination of leakage in weakly nonlinear qubits},
  author={Motzoi, F and Gambetta, J M and Rebentrost, P and Wilhelm, F K},
  journal={Phys. Rev. Lett.},
  volume={103},
  number={11},
  pages={110501},
  year={2009},
  publisher={APS},
  doi={10.1103/PhysRevLett.103.110501}
}

@article{gambetta2011analytic,
  title={Analytic control methods for high-fidelity unitary operations in a weakly nonlinear oscillator},
  author={Gambetta, J M and Motzoi, F and Merkel, S T and Wilhelm, F K},
  journal={Phys. Rev. A},
  volume={83},
  number={1},
  pages={012308},
  year={2011},
  publisher={APS},
  doi={10.1103/PhysRevA.83.012308}
}

@article{chen2016measuring,
  title={Measuring and suppressing quantum state leakage in a superconducting qubit},
  author={Chen, Zijun and Kelly, Julian and Quintana, Chris and Barends, Rami and Campbell, Brooks and Chen, Yu and Chiaro, Ben and Dunsworth, Andrew and Fowler, Austin G and Lucero, Erik and others},
  journal={Phys. Rev. Lett.},
  volume={116},
  number={2},
  pages={020501},
  year={2016},
  publisher={APS},
  doi={10.1103/PhysRevLett.116.020501}
}

@article{sheldon2016procedure,
  title={Procedure for systematically tuning up cross-talk in the cross-resonance gate},
  author={Sheldon, Sarah and Magesan, Easwar and Chow, Jerry M and Gambetta, Jay M},
  journal={Phys. Rev. A},
  volume={93},
  number={6},
  pages={060302},
  year={2016},
  publisher={APS},
  doi={10.1103/PhysRevA.93.060302}
}

@article{koch2007charge,
  title={Charge-insensitive qubit design derived from the {Cooper} pair box},
  author={Koch, Jens and Yu, Terri M and Gambetta, Jay and Houck, Andrew A and Schuster, David I and Majer, J and Blais, Alexandre and Devoret, Michel H and Girvin, Steven M and Schoelkopf, Robert J},
  journal={Phys. Rev. A},
  volume={76},
  number={4},
  pages={042319},
  year={2007},
  publisher={APS},
  doi={10.1103/PhysRevA.76.042319}
}

@article{lindblad1976generators,
  title={On the generators of quantum dynamical semigroups},
  author={Lindblad, G{\"o}ran},
  journal={Commun. Math. Phys.},
  volume={48},
  number={2},
  pages={119--130},
  year={1976},
  publisher={Springer},
  doi={10.1007/BF01608499}
}

@book{breuer2002theory,
  title={The Theory of Open Quantum Systems},
  author={Breuer, Heinz-Peter and Petruccione, Francesco},
  year={2002},
  publisher={Oxford University Press},
  doi={10.1093/acprof:oso/9780199213900.001.0001}
}

@article{wood2018quantification,
  title={Quantification and characterization of leakage errors},
  author={Wood, Christopher J and Gambetta, Jay M},
  journal={Phys. Rev. A},
  volume={97},
  number={3},
  pages={032306},
  year={2018},
  publisher={APS},
  doi={10.1103/PhysRevA.97.032306}
}

@article{norris2016qubit,
  title={Qubit noise spectroscopy for non-{Gaussian} dephasing environments},
  author={Norris, Leigh M and Paz-Silva, Gerardo A and Viola, Lorenza},
  journal={Phys. Rev. Lett.},
  volume={116},
  number={15},
  pages={150503},
  year={2016},
  publisher={APS},
  doi={10.1103/PhysRevLett.116.150503}
}

@article{preskill2018quantum,
  title={Quantum computing in the {NISQ} era and beyond},
  author={Preskill, John},
  journal={Quantum},
  volume={2},
  pages={79},
  year={2018},
  publisher={Verein zur F{\"o}rderung des Open Access Publizierens in den Quantenwissenschaften},
  doi={10.22331/q-2018-08-06-79}
}

@article{knill2005quantum,
  title={Quantum computing with realistically noisy devices},
  author={Knill, Emanuel},
  journal={Nature},
  volume={434},
  number={7029},
  pages={39--44},
  year={2005},
  publisher={Nature Publishing Group},
  doi={10.1038/nature03350}
}

@article{lucero2010reduced,
  title={Reduced phase error through optimized control of a superconducting qubit},
  author={Lucero, Erik and Kelly, Julian and Bialczak, Radoslaw C and Lenander, Mike and Mariantoni, Matteo and Neeley, Matthew and O'Connell, Aaron D and Sank, Daniel and Wang, Haohua and Weides, Martin and others},
  journal={Phys. Rev. A},
  volume={82},
  number={4},
  pages={042339},
  year={2010},
  publisher={APS},
  doi={10.1103/PhysRevA.82.042339}
}

@article{kelly2014optimal,
  title={Optimal quantum control using randomized benchmarking},
  author={Kelly, Julian and Barends, R and Campbell, B and Chen, Y and Chen, Z and Chiaro, B and Dunsworth, A and Fowler, A G and Hoi, I-C and Jeffrey, E and others},
  journal={Phys. Rev. Lett.},
  volume={112},
  number={24},
  pages={240504},
  year={2014},
  publisher={APS},
  doi={10.1103/PhysRevLett.112.240504}
}

@article{werninghaus2021leakage,
  title={Leakage reduction in fast superconducting qubit gates via optimal control},
  author={Werninghaus, Max and Egger, Daniel J and Roy, Farzad and Machnes, Shai and Wilhelm, Frank K and Filipp, Stefan},
  journal={npj Quantum Inf.},
  volume={7},
  number={1},
  pages={14},
  year={2021},
  publisher={Nature Publishing Group},
  doi={10.1038/s41534-020-00346-2}
}

@article{propson2022robust,
  title={Robust quantum optimal control with trajectory optimization},
  author={Propson, Thomas and Jackson, Brian E and Koch, Jens and Manchester, Zachary and Schuster, David I},
  journal={Phys. Rev. Appl.},
  volume={17},
  number={1},
  pages={014036},
  year={2022},
  publisher={APS},
  doi={10.1103/PhysRevApplied.17.014036}
}

@article{johansson2012qutip,
  title={{QuTiP}: An open-source {Python} framework for the dynamics of open quantum systems},
  author={Johansson, J Robert and Nation, Paul D and Nori, Franco},
  journal={Comput. Phys. Commun.},
  volume={183},
  number={8},
  pages={1760--1772},
  year={2012},
  publisher={Elsevier},
  doi={10.1016/j.cpc.2012.02.021}
}

@article{johansson2013qutip,
  title={{QuTiP} 2: A {Python} framework for the dynamics of open quantum systems},
  author={Johansson, J Robert and Nation, Paul D and Nori, Franco},
  journal={Comput. Phys. Commun.},
  volume={184},
  number={4},
  pages={1234--1240},
  year={2013},
  publisher={Elsevier},
  doi={10.1016/j.cpc.2012.11.019}
}

@misc{iqm2024garnet,
  title={{IQM} {Garnet}: A 20-qubit superconducting quantum processor},
  author={{IQM Quantum Computers}},
  year={2024},
  note={Technical specifications available at \url{https://www.meetiqm.com/}}
}

@article{krantz2019quantum,
  title={A quantum engineer's guide to superconducting qubits},
  author={Krantz, Philip and Kjaergaard, Morten and Yan, Fei and Orlando, Terry P and Gustavsson, Simon and Oliver, William D},
  journal={Appl. Phys. Rev.},
  volume={6},
  number={2},
  pages={021318},
  year={2019},
  publisher={AIP Publishing},
  doi={10.1063/1.5089550}
}

@article{arute2019quantum,
  title={Quantum supremacy using a programmable superconducting processor},
  author={Arute, Frank and Arya, Kunal and Babbush, Ryan and Bacon, Dave and Bardin, Joseph C and Barends, Rami and Biswas, Rupak and Boixo, Sergio and Brandao, Fernando GSL and Buell, David A and others},
  journal={Nature},
  volume={574},
  number={7779},
  pages={505--510},
  year={2019},
  publisher={Nature Publishing Group},
  doi={10.1038/s41586-019-1666-5}
}

@article{kjaergaard2020superconducting,
  title={Superconducting qubits: Current state of play},
  author={Kjaergaard, Morten and Schwartz, Mollie E and Braum{\"u}ller, Jochen and Krantz, Philip and Wang, Joel I-J and Gustavsson, Simon and Oliver, William D},
  journal={Annu. Rev. Condens. Matter Phys.},
  volume={11},
  pages={369--395},
  year={2020},
  publisher={Annual Reviews},
  doi={10.1146/annurev-conmatphys-031119-050605}
}

@article{blais2021circuit,
  title={Circuit quantum electrodynamics},
  author={Blais, Alexandre and Grimsmo, Arne L and Girvin, Steven M and Wallraff, Andreas},
  journal={Rev. Mod. Phys.},
  volume={93},
  number={2},
  pages={025005},
  year={2021},
  publisher={APS},
  doi={10.1103/RevModPhys.93.025005}
}

@misc{malarchick2026qubitpulseopt,
  author={Malarchick, Rylan},
  title={{QubitPulseOpt}: Quantum optimal control pulse optimization framework},
  year={2026},
  publisher={GitHub},
  url={https://github.com/rylanmalarchick/QubitPulseOpt},
  note={Includes simulation code, experiment scripts, and raw result data in the \texttt{results/} directory}
}

\end{document}